\documentclass[12pt,preprint]{aastex}

\newcommand\ha{H$\alpha$}
\newcommand\ca{\mbox{Ca\,{\sc ii}}}

\newcommand\fe{\mbox{Fe\,{\sc i}}}

\newcommand\mg{\mbox{Mg\,{\sc ii}}}
\newcommand\cii{\mbox{C\,{\sc ii}}}

\newcommand\siiv{\mbox{Si\,{\sc iv}}}
\newcommand{\kms}{\,km~s$^{-1}$}

\usepackage[normalem]{ulem}      
\usepackage{color}          

\usepackage{kotex}

\shorttitle{transient brightening in a light bridge}
\shortauthors{Song et al.}

\begin{document}


\title{Three-minute Sunspot Oscillations Driven by Magnetic Reconnection in a Light Bridge }


\author{Donguk Song\altaffilmark{1}, Jongchul Chae\altaffilmark{2}, Hannah Kwak\altaffilmark{2}, Ryouhei Kano\altaffilmark{1}, Vasyl Yurchyshyn\altaffilmark{3}, Yong-Jae Moon\altaffilmark{4}, Eun-Kyung Lim\altaffilmark{5}, and Jeongwoo Lee\altaffilmark{2}}

\email{donguk.song@nao.ac.jp}


\altaffiltext{1}{National Astronomical Observatory of Japan, National Institutes of Natural Sciences, 2-21-1 Osawa, Mitaka, Tokyo 181-8588, Japan}

\altaffiltext{2}{Astronomy Program, Department of Physics and Astronomy, Seoul National University, 1 Gwanak-ro, Gwanak-gu, Seoul 08826, Korea}

\altaffiltext{3}{Big Bear Solar Observatory, New Jersey Institute of Technology, 40386 North Shore Lane, Big Bear City, CA 92314-9672, USA}

\altaffiltext{4}{Department of Astronomy and Space Science, Kyung Hee University, Yongin 17104, Korea}

\altaffiltext{5}{Korea Astronomy and Space Science Institute 776, Daedeokdae-ro, Yuseong-gu, Daejeon 34055, Korea}


\begin{abstract}
We report a different type of three-minute chromospheric oscillations above a sunspot in association with a small-scale impulsive event in a light bridge. During our observations, we found a transient brightening in the light bridge. The brightening was composed of elementary bursts that may be a manifestation of fast repetitive magnetic reconnections in the light bridge. Interestingly, the oscillations in the nearby sunspot umbra were impulsively excited when the intensity of the brightening reached its peak. The initial period of the oscillations was about 2.3 minutes and then gradually increased to 3.0 minutes with time. In addition, we found that the amplitude of the excited oscillations was twice the amplitude of oscillations before the brightening. Based on our results, we propose that magnetic reconnection occurring in a light bridge can excite of oscillations in the nearby sunspot umbra. 


\end{abstract}


\keywords{Sun: UV radiation --- Sun: transition region --- Sun: chromosphere --- sunspots}

\section{INTRODUCTION}

Magnetohydrodynamic (MHD) waves and oscillations are ubiquitous phenomena in the solar atmosphere, especially above a sunspot. They have been regarded as one of the critical processes of energy transfer from the photosphere to the corona \citep{kanoh16}. The sunspot oscillations have been studied by analyzing the simultaneous time-series data with strong spectral lines formed in the different atmospheric layers, such as the \ha\ and \ca\ 8542 \AA\ lines \citep{jess12}, the \mg, \cii, and \siiv\ lines \citep{tian14}, the G-band and \ca\ K line \citep{krishna15}, and the \mbox{Na\,{\sc i} D2} and \fe\ 5434 \AA\ lines \citep[][]{chae16}. Even though there has been a conjecture that the external p-mode absorption and magnetoconvection can produce three-minute oscillations in a sunspot \citep[][]{khomenko15}, the origin of oscillations still remains one of the unresolved problems in the solar physics.

It has been suggested that turbulent convection in the photosphere can lead to the oscillations in a sunspot as well as the quiet Sun \citep{goldreich90}. One relevant observation is the finding of a good correlation of the amplitude modulation period between a sunspot and surrounding regions of the sunspot \citep{krishna15}. This may be also considered as a support of the theory that the p-mode absorption in the photosphere is a critical source of the sunspot oscillations \citep{cally94,rosenthal00}. On the other hand, the recent high-resolution observations of sunspots have revealed that three-minute oscillations occur inside the sunspot umbra in association with the umbral substructures such as light bridges (LBs) and umbral dots (UDs). \citet{yurchyshyn15} reported chromospheric umbral flashes that prefer to take place in the vicinities of LBs. Moreover, \citet{chae16} reported a local enhancement of three-minute oscillations power near a LB and UDs from the photospheric lines. These authors suggested that magnetoconvection occurring in LBs and UDs and non-unifomities of magnetic fields inside a sunspot umbra may excite the oscillations in a sunspot. 

Meanwhile, a different type of oscillations in a sunspot, which are generated by external impulsive events in the chromosphere, have been occasionally reported by several authors. Theoretically, \citet{kalkofen94} and \citet{chae15} suggested that a local disturbance generated by the impulsive events such as magnetic reconnection can lead to chromospheric oscillations of the acoustic cutoff frequencies in a disturbed region. In fact, \citet{kosovichev07} and \citet{kwak16} reported the observations of sunspot oscillations generated by impulsive events in the chromosphere: a solar flare and a strong downflow event, respectively. 

Recently, \citet{robustini16} and \citet{song16} reported jets in LBs that are thought to be generated by magnetic reconnection. A LB is a special region where a variety of dynamic phenomena such as chromospheric plasma ejections \citep{shimizu09} and transient brightening \citep{louis09} are frequently observed inside sunspot umbrae. It has been proposed that these kinds of events are caused by magnetic reconnection in the low chromosphere above a LB \citep{toriumi15,toriumi15b}. Here, we report an excitation of three-minute oscillations in the chromosphere above a sunspot umbra. In particular, we investigate a transient brightening in a LB and the associated oscillations in the nearby sunspot umbra by analyzing high-resolution spectral data taken by {\it Interface Region Imaging Spectrograph} \citep[IRIS;][]{pontieu14}. Our results provide a different type of three-minute chromospheric oscillations above a sunspot in association with the dynamics of the LB.

\section{OBSERVATIONS}
We analyzed a LB seen in a sunspot (X=8$''$,Y=-250$''$) of NOAA Active Region 12216 observed by IRIS from 02:12 UT to 03:12 UT on November 26, 2014. IRIS is a small explorer spacecraft that observes a wide range of wavelengths in ultraviolet spectral lines with high-temporal, spectral and spatial resolution. In particular, the IRIS can simultaneously record the near-ultraviolet (NUV, $2783-2834$~\AA) and far-ultraviolet (FUV, $1332-1358$~\AA\ and $1390-1406$~\AA) bands with the spectral dispersions being 0.025 \AA\,pixel$^{-1}$ in the NUV band and 0.013 \AA\,pixel$^{-1}$ in the FUV band.

We used the calibrated IRIS level 2 data obtained after dark current subtraction, flat-fielding, and geometrical correction \citep{pontieu14}. Our spectral data were taken in a large sit-and-stare observation whose cadence was 9 s. We mainly analyzed three strong emission spectra in the \mg~2796.3~\AA, \cii~1334.5~\AA, and \siiv~1402.8~\AA\ lines. The Doppler shift of each line was determined from applying the double-Gaussian fitting and the lambdameter method \citep{deubner96} to the \mg\ and \cii\ spectral lines at the middle wavelength of the chord ($\bigtriangleup\lambda \sim$ 0.02 \AA). 

In addition, we analized \mg\ 2796 \AA\ (chromosphere), \cii\ 1330 \AA\ (lower transtion region), and \siiv\ 1400 \AA\ (middle transition region) slit-jaw (SJ) images. The cadence of each SJ image is about 29 s and the FOV is 119$''$ $\times$ 119$''$. Photospheric images were acquired with the 
Helioseismic and Magnetic Imager \citep[HMI;][]{schou12} on board the Solar Dynamics Observatory \citep[SDO;][]{pesnell12}. The IRIS data were aligned with SDO/AIA data by using the cross-correlation between the IRIS \mg~2796~\AA\  SJ image and the SDO/AIA 1600 \AA\ image. The co-alignment between the SDO/AIA 1600 \AA\ image and the SDO/HMI intensity image was achieved by using the routines, `aia$\_$prep.pro' and `hmi$\_$prep.pro' available through the standard solar software pipeline, SSWIDL.

\section{RESULTS}

\begin{figure}
 \begin{center}
\includegraphics[width=1.\textwidth,clip=]{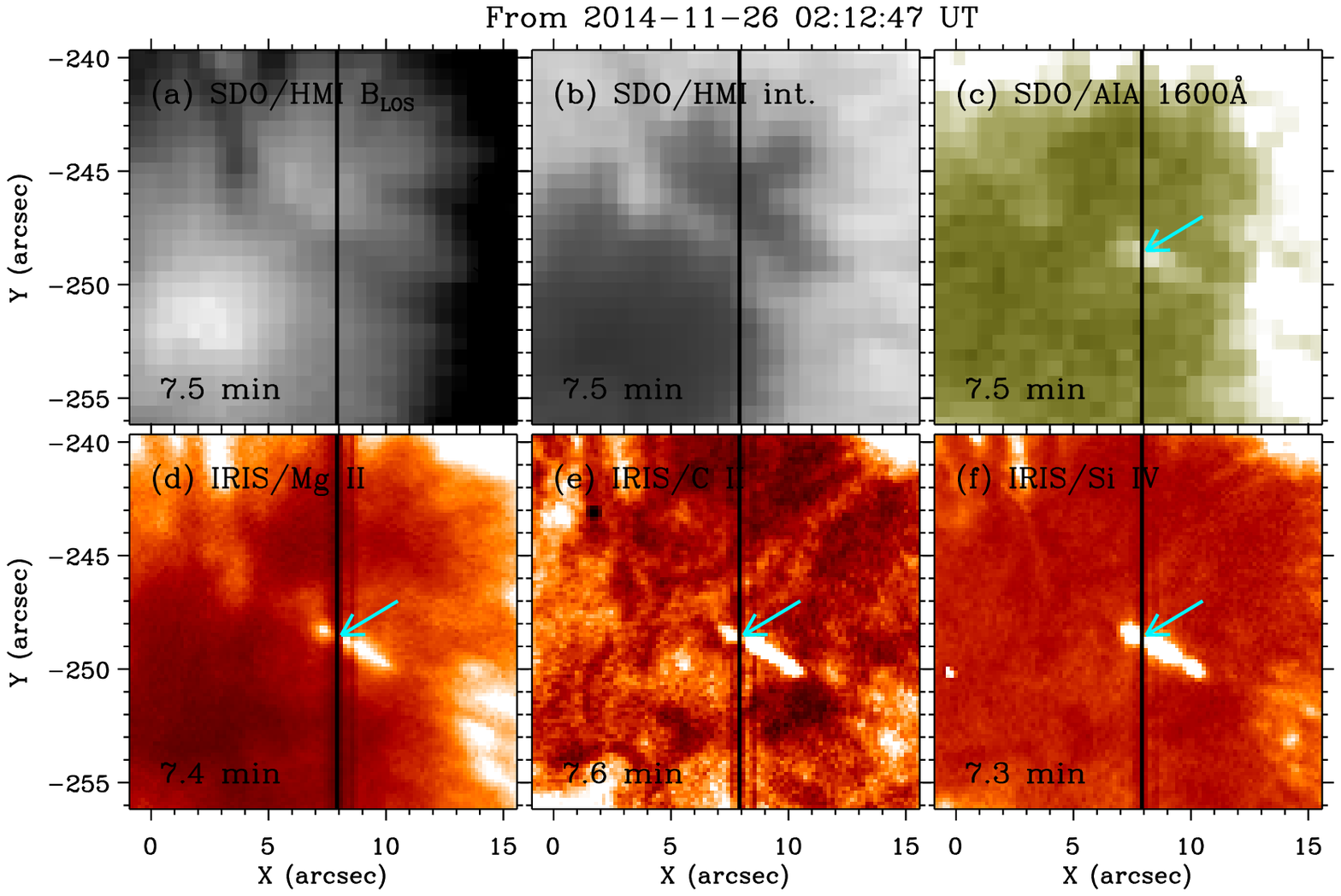}
\caption{Images of the region of our interest obtained on November 26, 2014: (a) the longitudinal magnetic field and (b) the continuum intensity of the SDO/HMI, (c) the SDO/AIA 1600 \AA\ intensity, (d) the \mg\, (e) \cii\, and (f) \siiv\ intensities of IRIS. The black solid lines present the position of the IRIS slit, and the cyan arrows indicate the bright feature occurring above the LB. (An animation of this figure is available.)   }\label{fig1}
 \end{center}
\end{figure}

Figure 1 shows the images of a part of the sunspot from the photosphere to the middle transition region. The sunspot with positive polarity has a filamentary LB dividing the sunspot umbra into two umbral regions. During our observations, we find a burtst of transient brightenings occurring along the LB (cyan arrows in Figure~\ref{fig1}). The brightenings repeatedly occurred at the same location with a short time interval of tens of seconds in all the IRIS SJ images of the \mg, \cii, and \siiv\ lines. 
 
\begin{figure}
 \begin{center}

 \epsscale{0.7}
\plotone{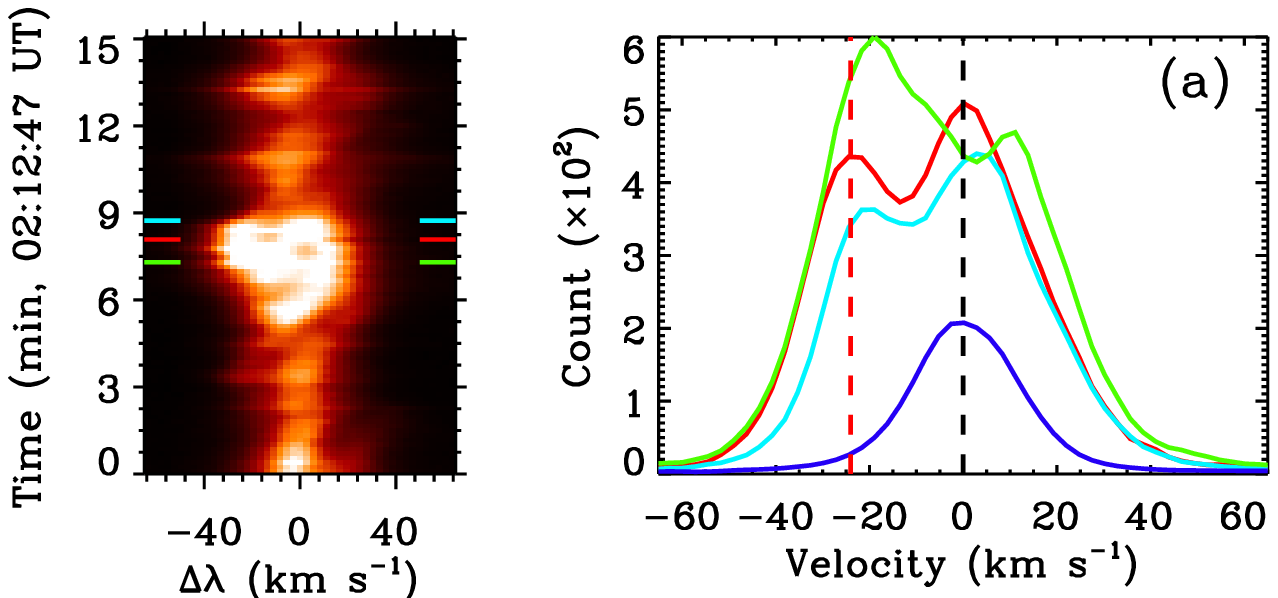}\\
\plotone{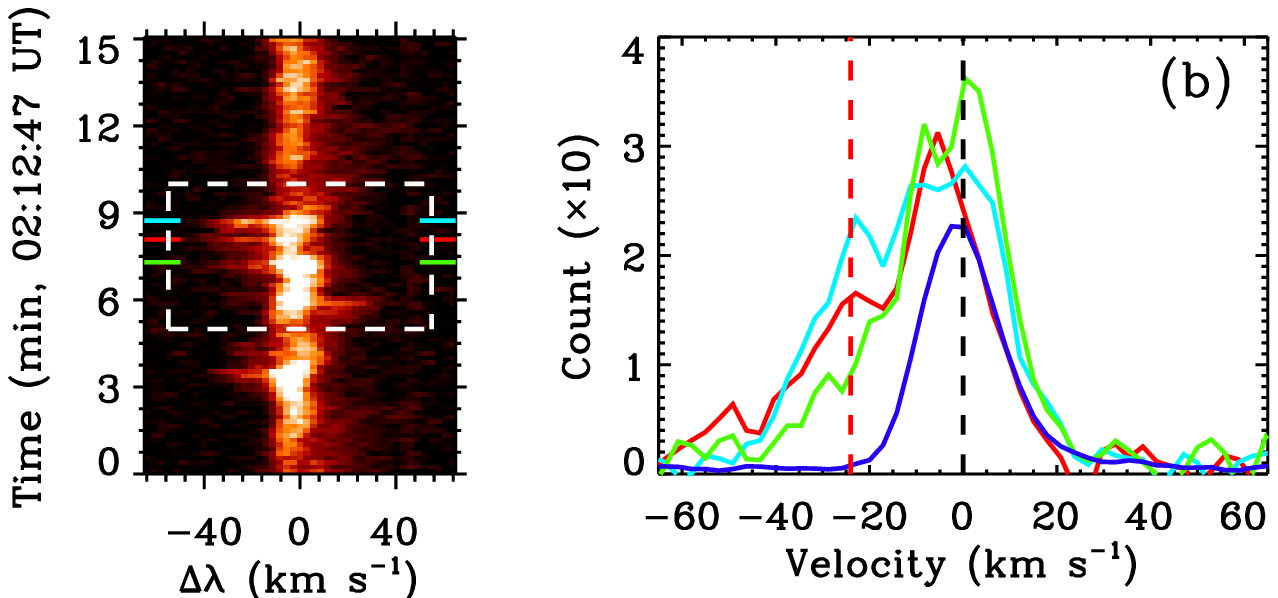}\\
\plotone{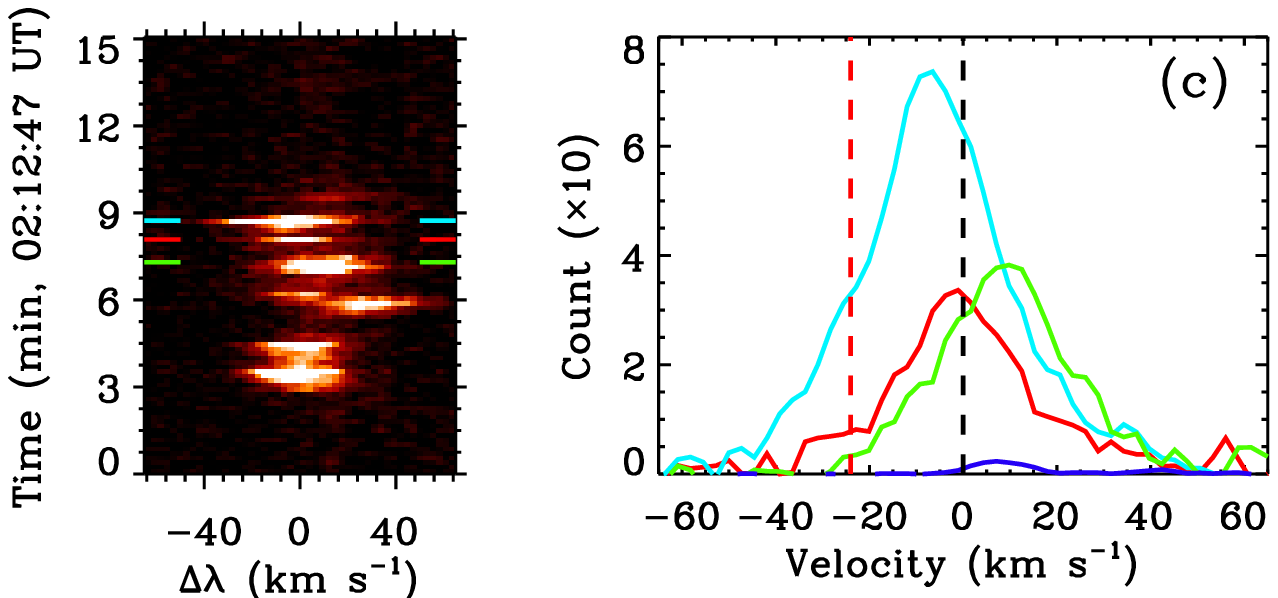}\\

\caption{Left : wavelength-time plots ($\lambda-t$ plots) of the \mg\ (top), \cii\ (middle), and \siiv\ (bottom) intensities at the LB. The horizontal lines colored in green, red, and cyan lines indicate the positions where we investigate the spectral profiles. The white dashed box encloses fine temporal intensity structures which we call elementary bursts. Right : spectral profiles of the enhanced brightenings in the (a) \mg, (b) \cii, and (c) \siiv\ lines. The blue solid lines represent the average profiles constructed from the spectral profiles of the LB during our observational time. Green, red, and cyan lines indicate the spectral profiles of the enhanced brightening in the LB, which follows the same color convention as the left panels. Red dashed lines mark the blue-shifted components of each line. }\label{fig2}
 \end{center}
\end{figure}

The transient brightenings are well identified in the $\lambda-t$ plots of the \mg, \cii, and \siiv\ intensity profiles (Figure~\ref{fig2}) from 02:18 UT ($t$ = 6 min) to 02:21UT ($t$ = 9 min). 
The transient brightenings consist of fine temporal structures, which are suggestive of elementary bursts \citep{jager78}. The time scale of each burst ranges from a few seconds to less than one minute (see the $\lambda-t$ plot of the \cii\ spectral profile). Each burst is  represented by broad emission spectral lines with enhanced intensity and high Doppler velocity. Note that the spectral profiles of the \mg\ and \cii\ lines have the shape of a double Gaussian, and the \siiv\ spectral profiles have the shape of a single Gaussian with the significantly enhanced line wings. These profiles are distinguished from the average profiles of all lines that present a sigle Gaussian (blue lines in Figure~\ref{fig2}).   
The core intensity of the transient brightening is enhanced by a factor of 2.9 in the \mg\ line and a factor of 1.6 in the \cii\ line compared to the core intensity of average profiles of each band. In addition, the intensity variations  at the blue wing are much larger than those at the core of the \mg\ and \cii\ lines by a factor of 2.0$-$4.0 times. This implies that plasma upflows may be predominant plasma motions during the transient brightening. The speed of the upwards motion determined from the double-Gaussian fit to the \mg\ and \cii\ lines was found to be about 24\kms. This value is comparable to the sound speed of either the chromosphere or the transition region, but is much  smaller than the Alfv{\'e}n speed typical for those regions.

\begin{figure}
 \begin{center}

 \epsscale{0.56}
\plotone{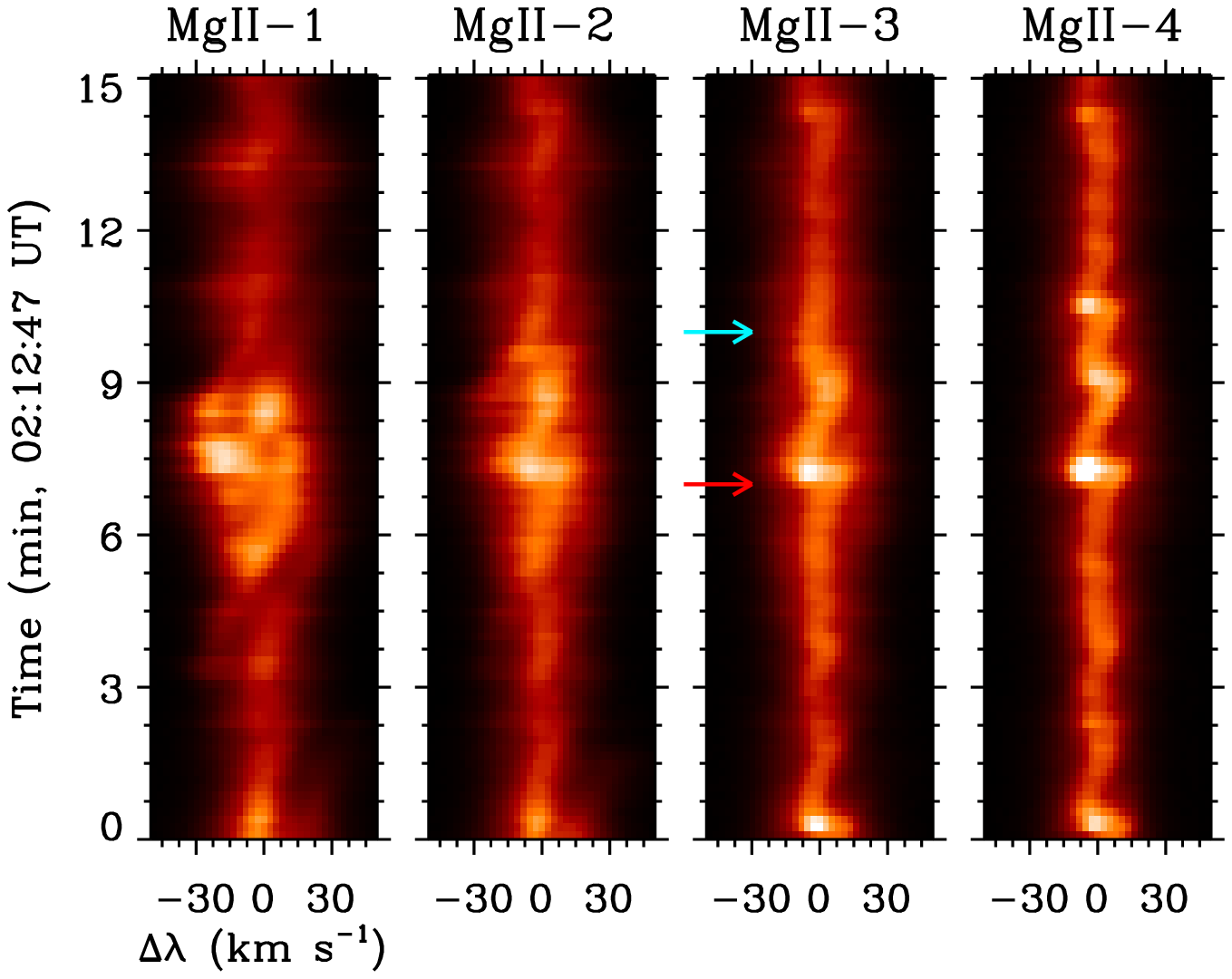}\\
\plotone{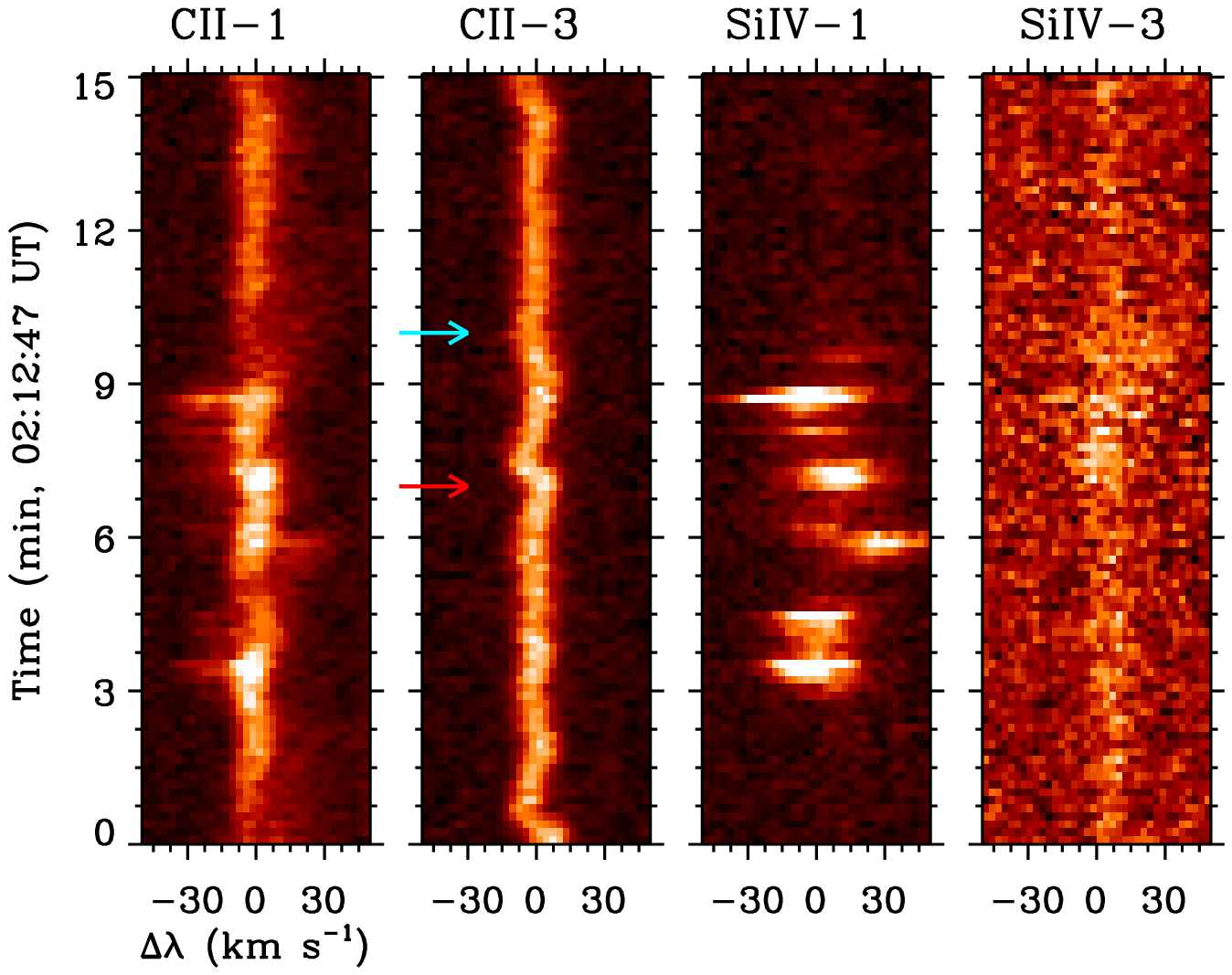}\\
 \epsscale{0.3}
\plotone{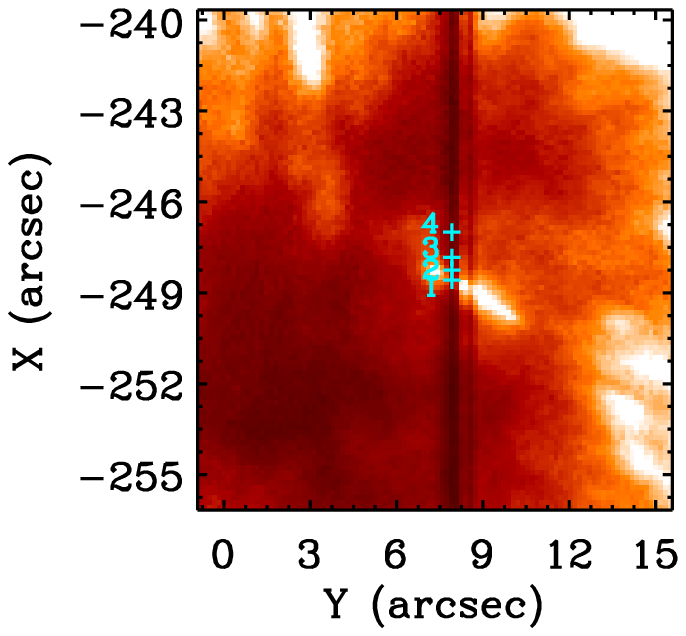}\\

\caption{Wavelength-time plots ($\lambda-t$ plots) of \mg, \cii, and \siiv\ intensities constructed at the different positions (plus marks in the bottom image) of the LB (position 1) and the sunspot umbra (position $2-4$). }\label{fig3}
 \end{center}
\end{figure}

One notable finding of the current study is that the neighboring region of the LB in the sunspot umbra was much affected by the transient brightening occurring above the LB. Figure~\ref{fig3} shows the $\lambda-t$ plots of the \mg, \cii, and \siiv\ intensity profiles measured at various positions above the LB (position 1) and the sunspot umbra (positions 2$-$4). 
In the $\lambda-t$ plots of the \mg\ intensity above the sunspot umbra (\mg-2$-$\mg-4), we do not detect any prominent and organized pattern in the Doppler shift prior to the bright enhancement above the LB ($t$ $<$ 6 min). However, we find that a sawtooth pattern (red arrow) indicative of a shock wave abruptly appeared when the brightening in the LB reached a significant level. A similar spectral behavior was also detected in the $\lambda-t$ plots of the \cii\ intensity profile above a sunspot umbra (\cii-3). At the same time we could not detect such a pattern in the $\lambda-t$ plot of the \siiv\ intensity profile, which may be due to the lower signal-to-noise of the \siiv\ line (\siiv-3). Meanwhile, we also found the bursts of transient brighteing in the \cii\ and \siiv\ lines (\cii-1 and \siiv-1) at $t$ $<$ 6 min, but such events above the LB did not trigger shock waves in the sunspot umbra. This implies that the two bright events occurring in the LB before and after 6 minutes have different properties. In particular, we note that the bright events seen in the $\lambda-t$ plots of the \cii\ and \siiv\ intensities at $t$ $<$ 6 min did not have their counterparts in the $\lambda-t$ plot of the \mg\ intensity. This indicates that such events took place in the transition region unlike the events of our interest. Thus, we think that the oscillations in the sunspot umbra are mainly driven by the impulsive events in the lower atmospheric layer.

\begin{figure}
\begin{center}
\epsscale{1.}
\plottwo{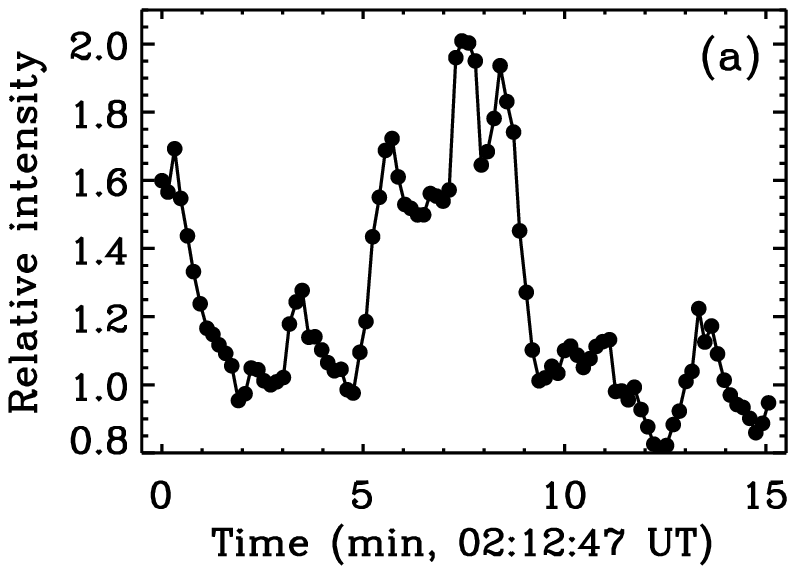}{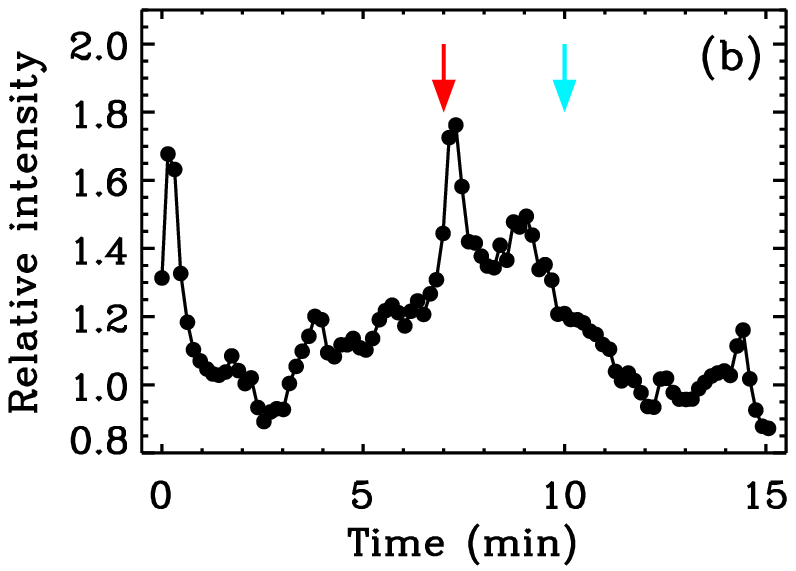}\\
\plottwo{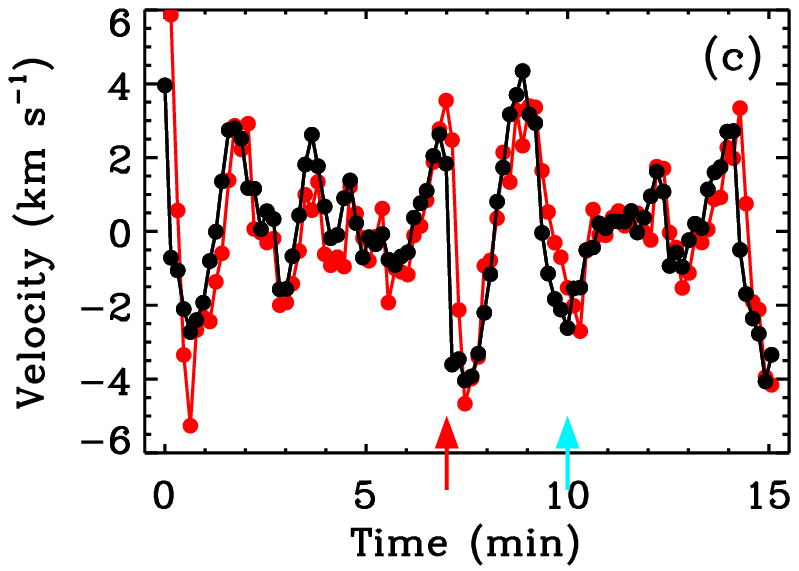}{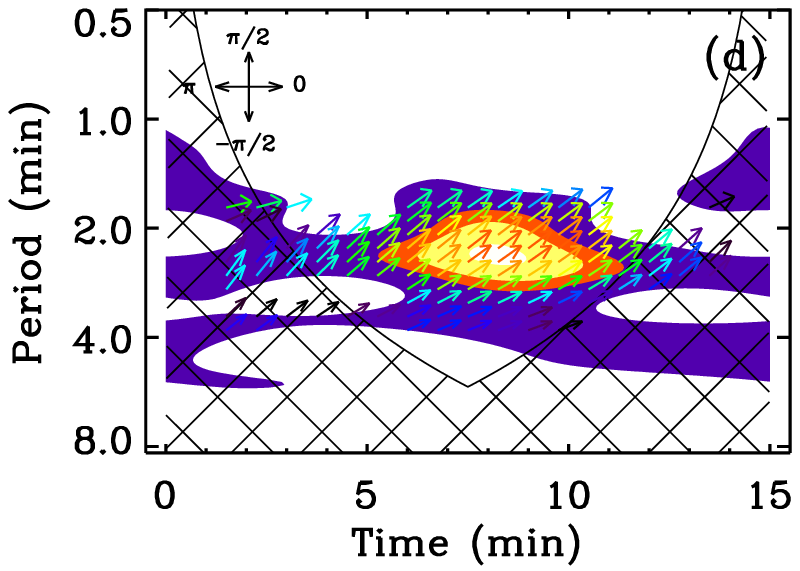}\\
\caption{Time series of the \mg\ intensity at the positions marked as (a) 1 and (b) 3 denoted in Figure 3. (c) Time variations of the line-of-sight Doppler velocity determined from the \mg\ (black line) and \cii\ (red line) bands. The red and cyan arrows indicate the same observational time as marked by the same color arrows in Figure 3. (d) The wavelet power spectrum of the \mg\ velocity. The direction of arrows indicate the phase difference of the velocities between the \mg\ and \cii\ lines, and the color of arrows indicates the coherency of the velocities: reddish arrows are close to the correlation value of 1, and bluish arrows are close to the correlation value of 0.9.}\label{fig4}
\end{center}
\end{figure}

The oscillatory behaviour above the sunspot umbra can be investigated in detail from time series of \mg\ and \cii\ intensities and velocities (Figure~\ref{fig4}). As mentioned above, the transient brightening first occured in the LB at around $t$ = 6 min and lasted  
 until around $t$ = 9 min with a peak value at around $t$ = 7 min (Figure~\ref{fig4}(a)). At this time, we also detected an enhancement of the \mg\ intensity by a factor of 1.5 at the position of the sunspot umbra (red arrow in Figure~\ref{fig4}(b)). The important point is that a sudden velocity jump (red arrow in Figure~\ref{fig4}(c)) from downflow with the speed of 2 \kms\ to upflow with the speed of 4 \kms\ appears when the brightening in the LB reaches its peak value. Note that, the sudden velocity jump is regarded as a manifestation of a shock front. 
 
 After the velocity jump, we find well-defined oscillation patterns. The amplitude of oscillations was about 4 \kms\ in the \mg\ and \cii\ lines, and dereased with time. From the wavelet power spectrum of the \mg\ velocity (Figure~\ref{fig4}(d)), we find several properties of the oscillations. First, its inital period was about 2.3 minutes and gradually increased to about 3.0 minutes. Second, the oscillation of the \cii\ velocity lagged the oscillation of the \mg\ velocity. This indicates that the waves we observed propagated upward from the chromosphere to the transition region since the two spectral lines form at different heights \citep{rathore15}. Finally, the coherency and the phase difference between the \mg\ and \cii\ velocity oscillations sharply change after the brightening event. In particular, the time lag of the velocity oscillations in these lines was about 19 s before the event that is consistent with the finding of \citet{tian14}, but the time lag change to about 9 s after the event. In addition, the coherence between the \mg\ and \cii\ velocities was strengthened after the occurrence of the brightening. This implies that a new wave phenomenon in a sunspot umbra may have been generated after the transient brightening event. 

\section{DISCUSSION}

We have reported a different type of three-minute oscillations above a sunspot in association with small-scale impulsive events 
in a LB. In particular, we provide an observational evidence that the event of transient brightenings in the LB impulsively leads to the chromospheric three-minute oscillations in the nearby sunspot umbra. 
This is consistent with the previous findings that a local enhancement of the three-minute sunspot oscillation power appears in the vicinities of LBs \citep{yurchyshyn15,chae16}. Meanwhile, the oscillations we detected are distinct from typical three-minute oscillation seen in a sunspot umbra for several reasons \citep{tian14}. First, the oscillations were impulsive and transient. Second, the initial period of the oscillations was 2.3 minutes and then gradually increased to 3.0 minutes. Third, the amplitude of oscillations was about two times larger than before occurring the brighenings, and was decreased with time. Our findings  correspond to  the characteristics of impulsively excited oscillations by a strong downflow event \citep{kwak16} and a solar flare \citep{kosovichev07}.


It is widely believed that the transient brighenings in a LB are energy release process by magnetic reconnection in the chromosphere and transition region \citep{berger03,louis15,toriumi15}. Interestingly, the brightenings we observed consisted of fine temporal structures, which are suggestive of elementary bursts. 
These kinds of bursts have been frequently reported in previous studies of impulsive events such as flares \citep{jager78,qiu06} and explosive events \citep[EEs;][]{gupta15}.  The elementary bursts are usually observed as short-period intensity variations ranging from a few seconds to less than one minute in the chromosphere and the transition region, and have been widely regarded as magnetic energy release occurring on small-scales.  \citet{chae98b} reported  recurrent EEs, as bursts, that are associated with repetitive fast magnetic reconnections in the transition region. Moreover, \citet{qiu06} reported elementary bursts in the \ha\ and X-ray flares, and suggested that such bursts may originate from the fragmented magnetic energy release.  In this regard, our results suggest a possibility that energy release by repetitive fast magnetic reconnections in a LB leads to the oscillations in the nearby sunspot.

Meanwhile, since the oscillations we observed were excited on the sunspot umbra which is a different region from the energy release site, 
there may be a concern about our suggestion that the impulsive events in a LB can lead to the chromospheric oscillations in a sunspot umbra. The previous studies of a LB by using the multi-wavelength data have revealed that the LB is a suitable region where it can persistently inject sufficient energy into a sunspot umbra \citep{hirzberger02,shimizu09,chae16}. For example, 
convective motion is not fully prohibited by the strong magnetic fields of the sunspot \citep{hirzberger02}. It has been suggested that the turbulent convection operating in the LB can  produce waves at frequencies above the acoustic cutoff in a sunspot umbra \citep{jacoutot08}. The other possible candidate for the injection of energy into a sunspot umbra is persistent and impulsive events in the LB such as chromospheric plasma ejections \citep{shimizu09,song16} and transient brightenings \citep{berger03,louis09,louis15}.  It has been proposed that these kinds of chromospheric phenomena originate from magnetic reconnection in the low chromosphere above the LB \citep{toriumi15,toriumi15b}, and play a significant role in heating of the solar atmosphere above a sunspot umbra \citep{yurchyshyn15,song16}. The energy release by repetitive magnetic reconnection in the LB can produce a local disturbance in the chromosphere not only itself but also the neighboring region such as a sunspot umbra. We conjecture that 
such a disturbance can lead to the chromospheric oscillations in a sunspot umbra with larger amplitudes and higher frequencies than before the brightening.  


In a way, our study support the idea that chromospheric disturbances generated by impulsive events such as magnetic reconnection lead to the chromospheric oscillations at the frequency of the acoustic cutoff \citep{kalkofen94, chae15}. \citet{chae15} theoretically reported that disturbances of a region in a gravitationally stratified medium can produce a wide range of frequency wave packets. The high-frequency waves, whose group speeds are as high as the sound speed, quickly escape from the disturbed region, while the waves with acoustic cutoff frequency linger for a long time in the region. They suggested that a series of impulsive disturbances can produce the persistent three-minute oscillations in the chromosphere. On the basis of our results, we propose that persistent impulsive events such as repetitive magnetic reconnections in a LB can lead to three-minute oscillations in the nearby sunspot. We expect that next generation solar telescopes with high resolution, such as the Daniel K. Inouye Solar Telescope, will help us to better understand the origin of sunspot oscillations associated to the dynamical behavior of LBs.

\acknowledgments
We greatly appreciate the referee's constructive comments. This work was partly supported by the National Research Foundation of Korea (NRF-2017R1A2B4004466). V.Y. acknowledges support from AFOSR FA9550-15-1-0322 and NSF AST-1614457 grants and KASI. We thank the international Space Science Institute in Bern for enabling interesting discussions. J.L. was supported by the 2017 Brainpool Program of the KOFST.



\end{document}